# A Generic Framework for Efficient and Effective Subsequence Retrieval

Haohan Zhu
Department of Computer Science
Boston University
zhu@cs.bu.edu

George Kollios
Department of Computer Science
Boston University
gkollios@cs.bu.edu

Vassilis Athitsos
Computer Science and Engineering Department
University of Texas at Arlington
athitsos@uta.edu

## ABSTRACT

This paper proposes a general framework for matching similar subsequences in both time series and string databases. The matching results are pairs of query subsequences and database subsequences. The framework finds all possible pairs of similar subsequences if the distance measure satisfies the "consistency" property, which is a property introduced in this paper. We show that most popular distance functions, such as the Euclidean distance, DTW, ERP, the Frechét distance for time series, and the Hamming distance and Levenshtein distance for strings, are all "consistent". We also propose a generic index structure for metric spaces named "reference net". The reference net occupies $O(n)$ space, where $n$ is the size of the dataset and is optimized to work well with our framework. The experiments demonstrate the ability of our method to improve retrieval performance when combined with diverse distance measures. The experiments also illustrate that the reference net scales well in terms of space overhead and query time.

## 1. INTRODUCTION

Sequence databases are used in many real-world applications to store diverse types of information, such as DNA and protein data, wireless sensor observations, music and video streams, and financial data. Similarity-based search in such databases is an important functionality, that allows identifying, within large amounts of data, the few sequences that contain useful information for a specific task at hand. For example, identifying the most similar database matches for a query sequence can be useful for classification, forecasting, or retrieval of similar past events.

The most straightforward way to compare the similarity between two sequences is to use a global similarity measure, that computes an alignment matching the entire first sequence to the entire second sequence. However, in many scenarios it is desirable to perform *subsequence matching*, where, given two sequences $Q$ and $X$, we want to identify pairs of subsequences $SQ$ of $Q$ and $SX$ of $X$, such that the similarity between $SQ$ and $SX$ is high. When a large database of sequences is available, it is important to be able to identify, given a query $Q$, an optimally matching pair $SQ$ and $SX$, where $SX$ can be a subsequence of any database sequence.

A well-known example of the need for subsequence matching is in comparisons of biological sequences. It is quite possible that two DNA sequences $Q$ and $X$ have a large Levenshtein distance [22] (also known as edit distance) between them (e.g., a distance equal to 90% of the length of the sequences), while nonetheless containing subsequences $SQ$ and $SX$ that match at a very high level of statistical significance. Identifying these optimally matching subsequences [34] helps biologists reason about the evolutionary relationship between $Q$ and $X$, and possible similarities of functionality between those two pieces of genetic code.

Similarly, subsequence matching can be useful in searching music databases, video databases, or databases of events and activities represented as time series. In all the above cases, while the entire query sequence may not have a good match in the database, there can be highly informative and statistically significant matches between subsequences of the query and subsequences of database sequences.

Several methods have been proposed for efficient subsequence matching in large sequence databases. However, all the proposed techniques are targeted to specific distance or similarity functions, and it is not clear how and when these techniques can be generalized and applied to other distances. Especially, subsequence retrieval methods for string databases are difficult to be used for time-series databases. Furthermore, when a new distance function is proposed, we need to develop new techniques for efficient subsequence matching. In this paper we present a general framework, which can be applied to any arbitrary distance metric, as long as the metric satisfies a specific property that we call "consistency". Furthermore, we show that many well-known existing distance functions satisfy consistency. Thus, our framework can deal with both sequence types, i.e., strings and time series, including cases where each element of the sequence is a complex object.

The framework in this paper consists of a number of steps: dataset segmentation, query segmentation, range query, candidate generation, and subsequence retrieval. Brute-force search would require evaluating a total of $O\left(|Q|^2 |X|^2\right)$ pairs of subsequences of $Q$ and $X$. However our filtering method produces a shortlist of candidates after considering $O\left(|Q| |X|\right)$ pairs of segments only. For the case where the distance is a metric, we also present a hierarchical reference net, a novel generic index structure that can be used within our framework to provide efficient query processing.

Overall, this paper makes the following contributions:

- We propose a framework that, compared to alternative methods, makes minimal assumptions about the underlying distance, and thus can be applied to a large variety of distance functions.

- We introduce the notion of "consistency" as an important property for distance measures applied to sequences.

- We propose an efficient filtering method, which produces a shortlist of candidates by matching only $O(|Q| |X|)$ pairs of subsequences, whereas brute force would match $O(|Q|^2 |X|^2)$ pairs of subsequences.

- We make this filtering method even faster, by using a generic indexing structure with linear space based on reference nets, that efficiently supports range similarity queries.

- Experiments demonstrate the ability of our method to provide good performance when combined with diverse metrics such as the Levenshtein distance for strings, and ERP [8] and the discrete Frechét distance [11]) for time series.

## 2. RELATED WORK

Typically, the term "sequences" can refer to two different data types: strings and time-series. There has been a lot of work in subsequence retrieval for both time series and string databases. However, in almost all cases, existing methods concentrate on a specific distance function or specific type of queries. Here we review some of the recent works on subsequence matching. Notice that this review is not exhaustive since the topic has received a lot of attention and a complete survey is beyond the scope of this paper.

Time-series databases and efficient similarity retrieval have received a lot of attention in the last two decades. The first method for subsequence similarity retrieval under the Euclidean ($L_2$−norm) distance appeared in the seminal paper of Faloutsos et al. [12]. The main idea is to use a sliding window to create smaller sequences of fixed length and then use a dimensionality reduction technique to map each window to a small number of features that are indexed using a spatial index (e.g., $R^*$-tree). Improvements of this technique have appeared in [28, 27] that improve both the window-based index construction and the query time using sliding windows on the query and not on the database. However, all these techniques are applicable to the Euclidean distance only.

Another popular distance function for time series is the Dynamic Time Warping (DTW) distance [4, 17]. Subsequence similarity retrieval under DTW has also received some interest recently. An approach to improve the dynamic programming algorithm to compute subsequence matchings under DTW for a single streaming time series appeared in [32]. In [14] Han et al. proposed a technique that extends the work by Keogh et al. [16] to deal with subsequence retrieval under DTW. An improvement over this technique appeared in [15]. An efficient *approximate* subsequence matching under DTW was proposed in [3] that used reference based indexing for efficient retrieval. Another technique that uses early abandoning during the execution of dynamic programming appeared in [2] and a work that provides very fast query times even for extremely large datasets appeared recently [31]. An interesting new direction is to use FPGAs and GPUs for fast subsequence matching under DTW [33]. Again, all these works are tailored to the DTW distance.

Subsequence retrieval for string datasets has also received a lot of attention, especially in the context of biological data like DNA and protein sequences. BLAST [1] is the most popular tool that is used in the bioinformatics community for sequence and subsequence alignment of DNA and protein sequences. However, it has a number of limitations, including the fact that it is a heuristic and therefore may not report the optimal alignment. The distance functions that BLAST tries to approximate are variations of the Edit distance, with appropriate weights for biological sequences [34, 29]. A number of recent works have proposed new methods that improve the quality of the results and the query performance for large query sizes. RBSA is an embedding based method that appeared in [30] and that works well for large queries on DNA sequences. BWT-SW [20] employs a suffix array and efficient compression to speed up local alignment search. Other techniques target more specialized query models. A recent example is WHAM [25] that uses hash based indexing and bit-wise operations to perform efficient alignment of short sequences. Other techniques for similar problems include [21, 24, 13]. Finally, a number of techniques that use $q$-grams [38, 23, 18, 26] have been proposed for string databases, and can also be applied to biological datasets. However, all of these methods are applicable to specific data types and query models.

Indexing in metric spaces has been studied a lot in the past decades and a number of methods have been proposed. Tree-based methods include the Metric-tree [9] for external memory and the Cover-tree [6] for main memory. The cover-tree is a structure that provides efficient and *provable* logarithmic nearest neighbor retrieval under specific and reasonable assumptions about the data distribution. Other tree-based structures for metric spaces include the vp-tree [39] and the mvp-tree [7]. A nice survey on efficient and provable index methods in metric spaces is in [10].

Another popular approach is to use a set of carefully selected references and pre-compute the distance of all data in the database against these references [36]. Given a query, the distance of the references is computed first and using the triangular inequality of the metric distance, parts of the database can be pruned without computing all the distances. One problem with this approach is the large space requirement in practice.

## 3. PRELIMINARIES

Here we give the basic concepts and definitions that we use to develop our framework. Let $Q$ be a query sequence with length $|Q|$, and $X$ be a database sequence with length $|X|$. $Q = (q_1, q_2, q_3, ..., q_{|Q|})$ and $X = (x_1, x_2, x_3, ..., x_{|X|})$, where the individual values $q_i$, $x_j$ are elements of an alphabet $\Sigma$. In string databases, $\Sigma$ is a finite set of characters. For example: in DNA sequences, $\Sigma_D = \{A, C, G, T\}$, $|\Sigma_D| = 4$, while in protein sequences, $\Sigma_P$ contains 20 letters. The alphabet $\Sigma$ can also be a multi-dimensional space and/or an infinite set, which is the typical case if $Q$ and $X$ are time series. For example: for trajectories on the surface of the Earth, $\Sigma_T = \{(longitude, latitude)\} \subseteq \mathbb{R}^2$, $|\Sigma_T| = \infty$. Similarly, in tracks over a 3D space, $\Sigma_T = \{(x, y, z)\} \subseteq \mathbb{R}^3$, $|\Sigma_T| = \infty$.

For sequences defined over an alphabet $\Sigma_\phi$, we can choose a distance measure $\delta_\psi$ to measure the dissimilarity between any two sequences. We say that sequence $Q \in (\Sigma_\phi, \delta_\psi)$ when we want to explicitly specify the alphabet and distance measure employed in a particular domain.

## 3.1 Similar Subsequences

Let $SX$ with length $|SX|$ be a subsequence of $X$, and $SQ$ with length $|SQ|$ be a subsequence of $Q$. We denote $SX_{a,b}$ as a subsequence with elements $(x_a, x_{a+1}, x_{a+2}, ..., x_b)$, and the elements of $SQ_{c,d}$ are $(q_c, q_{c+1}, q_{c+2}, ..., q_d)$. Subsequence $SX$ and $SQ$ should be continuous. Determining whether $SX$ and $SQ$ are similar will depend on two parameters, $\varepsilon$ and $\lambda$. Namely, we define that $SX$ and $SQ$ are similar to each other if the distance $\delta(SX, SQ)$ is not larger than $\varepsilon$, and the lengths of both $SX$ and $SQ$ are not less than $\lambda$.

Setting a minimal length parameter $\lambda$ allows us to discard certain subsequence retrieval results that are not very meaningful. For example:

- Two subsequences $SX$ and $SQ$ can have very small distance (even distance 0) to each other, but be too small (e.g., of length 1) for this small distance to have any significance. In many applications, it is not useful to consider such subsequences as meaningful matches. For example: two DNA sequences $X$ and $Q$ can each have a length of one million letters, but contain subsequences $SX$ and $SQ$ of length 3 that are identical.

- If the distance allows time shifting (like DTW [16], ERP [8], or the discrete Frechét distance [11]), a long subsequence can have a very short distance to a very short subsequence. For example: sequence 111222333 according to DTW has a distance of 0 to sequence 123. Using a sufficiently high $\lambda$ value, we can prevent the short sequence to be presented to the user as a meaningful match for the longer sequence.

As a matter of fact, later in the paper we will also use an additional parameter $\lambda_0$, that explicitly restricts the time shifting that is allowed between similar subsequences. The difference between the lengths of $SX$ and $SQ$ should not be larger than $\lambda_0$.

## 3.2 Query Types

Given a query sequence $Q$, there are three types of subsequence searches that we consider in this paper:

- Type I, Range Query: Return all pairs of similar subsequences $SX$ and $SQ$, where $|SX| \geqslant \lambda$, $|SQ| \geqslant \lambda$, $||SX| - |SQ|| \leqslant \lambda_0$ and $\delta(SX, SQ) \leqslant \varepsilon$.

  We should note that in typical cases, when $SX$ and $SQ$ are long and similar to each other, any subsequence of $SX$ has a similar subsequence in $SQ$. This observation is formalized in Section 4, using our definition of "consistency". In such cases, the search may return a large number of results that are quite related.

- Type II, Longest Similar Subsequence Query: Maximize $|SQ|$, subject to $|SX| \geqslant \lambda$, $||SX| - |SQ|| \leqslant \lambda_0$ and $\delta(SX, SQ) \leqslant \varepsilon$.

- Type III, Nearest Neighbor Query: Minimize $\delta(SX, SQ)$, subject to $|SX| \geqslant \lambda$, $|SQ| \geqslant \lambda$ and $||SX| - |SQ|| \leqslant \lambda_0$.

Since the first query type may lead to too many related results, the second and third query types are more practical. In Section 7, we introduce methods to retrieve query results after we generate similar segment candidates.

In this paper, we allow $\varepsilon$ to vary at runtime, whereas we assume that $\lambda$ is a user-specified parameter. Our rational is that, for a specific application, specific values of $\lambda$ may make sense, such as one hour, one day, one year, one paragraph, or values related to a specific error model. Thus, the system can be based on a predefined value of $\lambda$ that is appropriate for the specific data. On the other hand, $\varepsilon$ should be allowed to change according to different queries.

## 3.3 Using Metric Properties for Pruning

If distance $\delta$ is metric, then $\delta$ is symmetric and has to obey the triangle inequality, so that $\delta(Q, R) + \delta(R, X) \geqslant \delta(Q, X)$ and $\delta(Q, R) - \delta(R, X) \leqslant \delta(Q, X)$. The triangle inequality can be used to reject, given a query sequence $Q$, candidate database matches without evaluating the actual distance between $Q$ and those candidates.

For example, suppose that $R$ is a preselected database sequence, and that $r$ is some positive real number. Further suppose that, for a certain set $L$ of database sequences, we have verified that $\delta(R, X) > r$ for all $X \in L$. Then, given query sequence $Q$, if $\delta(Q, R) \leqslant r - \varepsilon$, the triangle inequality guarantees that, for all $X \in L$, $\delta(Q, X) > \varepsilon$. Thus, the entire set $X$ can be discarded from further consideration.

This type of pruning can only be applied when the underlying distance measure obeys the triangle inequality. Examples of such distances are the Euclidean distance, ERP, or the Frechét distance for time series, as well as the Hamming distance or Levenshtein distance for strings. DTW, on the other hand, does not obey the triangle inequality.

## 4. THE CONSISTENCY PROPERTY

As before, let $Q$ and $X$ be two sequences, and let $SQ$ and $SX$ be respectively subsequences of $Q$ and $X$. A simple way to identify similar subsequences $SQ$ and $SX$ would be to exhaustively compare every subsequence of $Q$ with every subsequence of $X$. However, that brute-force approach would be too time consuming.

In our method, as explained in later sections, we speed up subsequence searches by first identifying matches between segments of the query and each database sequence. It is thus important to be able to reason about distances between subsequences of $SQ$ and $SX$, in the case where similar subsequences $SQ$ and $SX$ do exist. As we want our method to apply to a more general family of distance measures, we need to specify what properties these measures must obey in order for our analysis to be applicable. One property that our analysis requires is a notion that we introduce in this paper, and that we term "consistency". Furthermore, if the distance function satisfies the triangular inequality, we can use efficient indexing techniques to improve the query time. However, we want to point out, that some distances with the consistency property may violate triangular inequality or symmetry. The consistency property is defined as follows:

*Definition 1.* We call distance $\delta$ a *consistent* distance measure if it obeys the following property: if $Q$ and $X$ are two sequences, then for every subsequence $SX$ of $X$ there exists a subsequence $SQ$ of $Q$ such that $\delta(SQ, SX) \leqslant \delta(Q, X)$.

At a high level, a consistent distance measure guarantees that, if $Q$ and $X$ are similar, then for every subsequence of $Q$ there exists a similar subsequence in $X$. From the definition of consistency, we can derive a straightforward lemma that can be used for pruning dissimilar subsequences:

*Lemma 1.* If the distance $\delta$ is consistent and $\delta(Q, X) \leqslant \varepsilon$, then for any subsequence $SX$ of $X$ there exists a subsequence $SQ$ of $Q$ such that $\delta(SQ_{j,k}, SX) \leqslant \varepsilon$.

We will show in the next paragraphs that the consistency property is satisfied by the Euclidean distance, the Hamming distance, the discrete Frechét distance, DTW, ERP, and the Levenshtein distance (edit distance).

We start with the Euclidean distance, which is defined as $\delta_E(Q, X) = (\sum_{m=1}^{d}(Q_m - X_m)^2)^{1/2}$, where $|Q| = |X| = d$. Then, for any subsequence $SQ_{ij}$ in $Q$, there exists a subsequence $SX_{ij}$ in $X$, such that $\delta_E(SQ, SX) = (\sum_{m=i}^{j}(Q_m - X_m)^2)^{1/2}$. Obviously, $\delta_E(SQ, SX)$ sums up only a subset of the terms that $\delta_E(Q, X)$ sums up, and thus $\delta_E(SQ, SX) \leqslant \delta_E(Q, X)$. Therefore, the Euclidean distance is consistent. The same approach can be used to show that the Hamming distance is also consistent.

Although DTW, the discrete Frechét distance, ERP, and the Levenshtein distance allow time shifting or gaps, they can also be shown to be consistent. Those distances are computed using dynamic programming algorithms that identify, given $X$ and $Q$, an optimal alignment $C$ between $X$ and $Q$. This alignment $C$ is expressed as a sequence of couplings: $C = (\omega_k, 1 \leqslant k \leqslant K)$, where $K \leqslant |X| + |Q|$, and where each $\omega_k = (i, j)$ specifies that element $x_i$ of $X$ is coupled with element $q_j$ of $Q$.

In an alignment $C$, each coupling incurs a specific cost. This cost can only depend on the two elements matched in the coupling, and possibly on the preceding coupling as well. For example, in DTW the cost of a coupling is typically the Euclidean distance between the two time series elements. In ERP and the edit distance, the cost of a coupling also depends on whether one of the two elements of the coupling also appears in the previous coupling.

While DTW, ERP, and the Levenshtein distance assign different costs to each coupling, they all define the optimal alignment $C$ to be the one that minimizes the sum of costs of all couplings in $C$. The discrete Frechét distance, on the other hand, defines the optimal alignment to be the one that minimizes the maximal value of its couplings.

For all four distance measures, the alignment $C$ has to satisfy certain properties, namely boundary conditions, monotonicity, and continuity [16]. Now, suppose that $SX_{a,b} = (x_a, x_{a+1}, x_{a+2}, ..., x_b)$ is a subsequence of $X$. For any element $x_i$ of $SX$ there exists some corresponding elements $q_j$ of $Q$ such that $(x_i, q_j)$ is a coupling $\omega_{(i,j)}$ in $C$. Suppose that the earliest matching element for $x_a$ is $q_c$ and the last matching element of $x_b$ is $q_d$, and define $SQ_{c,d} = (q_c, q_{c+1}, q_{c+2}, ..., q_d)$. Because of the continuity and monotonicity properties, $SQ_{c,d}$ is a subsequence of $Q$. Furthermore, the sequence of couplings in $C$ which match an element in $SX$ with an element in $SQ$ form a subsequence $SC$ of $C$, and $SC$ is an optimal alignment between $SX$ and $SQ$.

It follows readily that the sum or maximum of the subsequence $SC$ cannot be larger than the sum or maximum of the whole sequence $C$. Namely, $\delta(SX, SQ) \leqslant \delta(X, Q)$. This shows that $\forall\ SX_{a,b}, \exists\ SQ_{c,d}$, such $\delta(SX_{a,b}, SQ_{c,d}) \leqslant \delta(X, Q)$. Thus DTW, the discrete Frechét distance, ERP, and the Levenshtein distance are all "consistent".

## 5. SEGMENTATION

Let $X = (x_1, x_2, x_3, ..., x_{|X|})$ and $Q = (q_1, q_2, q_3, ..., q_{|Q|})$ be two sequences. We would like to find a pair of subsequences $SX_{a,b} = (x_a, x_{a+1}, x_{a+2}, ..., x_b)$ and $SQ_{c,d} = (q_c, q_{c+1}, q_{c+2}, ..., q_d)$, such that $\delta(SX, SQ) \leqslant \varepsilon$ and $|SX| \geqslant \lambda$, $|SQ| \geqslant \lambda$. Brute force search would check all $(a, c)$ combinations and all possible $|SX|$ and $|SQ|$. However, there are $O(|Q|^2)$ different subsequences of $Q$, and $O(|X|^2)$ different subsequences of $X$, so brute force search would need to evaluate $O(|Q|^2|X|^2)$ potential pairs of similar subsequences. The resulting computational complexity is impractical.

With respect to subsequences of $X$, we can drastically reduce the number of subsequences that we need to evaluate, by partitioning sequence $X$ into fixed-length windows $w_i$ with length $l$, so that $X = (w_1, w_2, w_3, ..., w_{|W|})$, where $w_i = (x_{(i-1)*l+1}, x_{(i-1)*l+2}, ..., x_{i*l})$, $(i \leqslant |Q|/l)$. The following lemma shows that matching segments of $Q$ only against such fixed-length windows can be used to identify possible locations of all subsequence matches:

*Lemma 2.* Let $SX$ and $SQ$ be subsequences with lengths $\geqslant \lambda$, such that $\delta(SQ, SX) \leqslant \varepsilon$, where $\delta$ is consistent. Let sequence $X$ be partitioned into windows of fixed length $l$. If $l \leqslant \lambda/2$, then there exists a window $w_j$ in $SX$ and a subsequence $SSQ$ from $SQ$, such that $\delta(SSQ, w_j) \leqslant \varepsilon$.

Proof: If the length of the fixed-length windows is less than or equal to $\lambda/2$, then for any subsequence $SX$ of $X$ with length at least $\lambda$, there must be a window $w_i$ that is fully contained within $SX$. Since $\delta$ is consistent, and based on lemma 1, if there exist subsequences $SQ$ and $SX$ such that $\delta(SQ, SX) \leqslant \varepsilon$, then if $w_i$ is a sub-subsequence of $SX$, there must be some sub-subsequence $SSQ$ from $SQ$ such that $\delta(SSQ, w_i) \leqslant \varepsilon$.

Based on lemma 2 we can obtain another straightforward lemma, that can be used for pruning dissimilar subsequences:

*Lemma 3.* Let sequence $X$ be partitioned into fixed-length windows with length $l = \lambda/2$, and let distance $\delta$ be consistent. If there is no subsequence $SQ$ such that $\delta(SQ, w_j) \leqslant \varepsilon$, then all subsequences which cover window $w_j$ have no similar subsequence in $Q$ (where "similar" is defined as having distance at most $\varepsilon$ and length at least $\lambda$).

To find a pair of similar subsequences between two sequences, we could partition one sequence into fixed-length windows and match those windows with sliding windows of the second sequence. Since the total length of sequences $X$ in the database is much larger than the length of query sequence $Q$, we partition sequences $X$ in the database into fixed-length windows with length $\lambda/2$, whereas from the query $Q$ we extract (using sliding windows) subsequences with different lengths.

If we use brute force search, there are a total of $O(|Q|^2 |X|^2)$ potential pairs of similar subsequences that need to be checked. When we partition sequences $X$ in the database as described above, there are only $(|X|/l)$ windows that need to be compared to query subsequences. Thus, the number of subsequence comparisons involved here is $O(|Q|^2|X|)$.

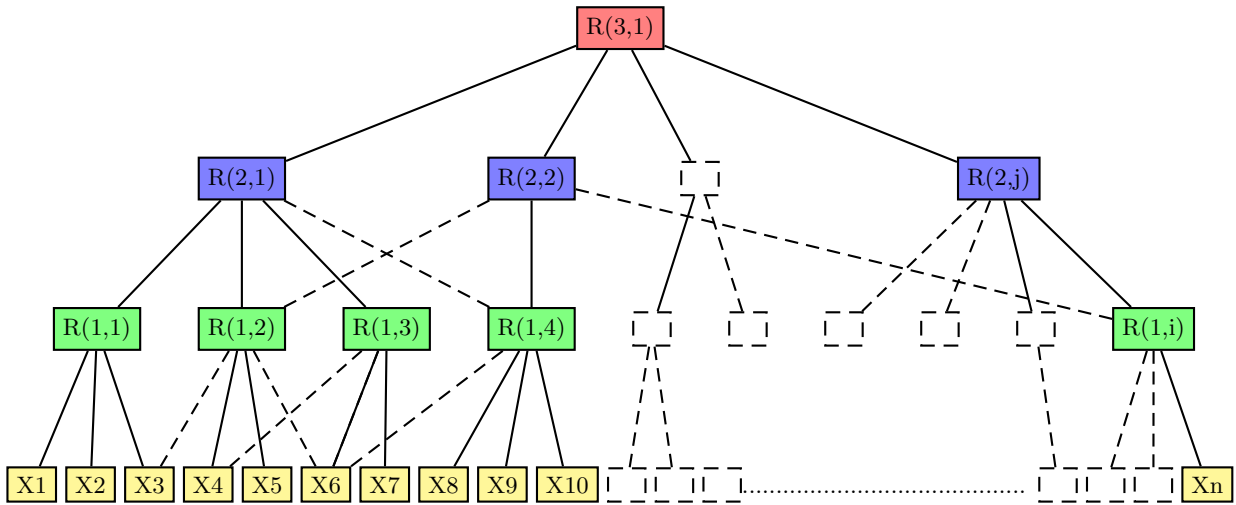

Figure 1: An example of a Reference Net

For a query sequence $Q$, there are about $(|Q|^2/2)$ different subsequences of $Q$ with length at least $\lambda$. However, we can further reduce the number of subsequence comparisons if we limit the maximum temporal shift that can occur between similar subsequences. In particular, if $\lambda_0$ is the maximal shift that we permit, then there are no more than $(2\lambda_0 + 1)|Q|$ different segments of $Q$ need to be considered. The total number of pairs of segments is no more than $2(2\lambda_0 + 1)|X||Q|/\lambda$. If $\lambda_0 \ll \lambda$, the number of pairs of segments is much less than $|X||Q|$.

The method that we propose in the next sections assumes both metricity and consistency. Thus, DTW is not suitable for our method, as it is not metric. While the Euclidean distance and the Hamming distance satisfy both metricity and consistency, they cannot tolerate even the slightest temporal misalignment or a single gap, and furthermore they require matches to always have the same length. These two limitations make the Euclidean and Hamming distances not well matched for sequence matching. Meanwhile, the discrete Fréchet distance, the ERP distance, and the Levenshtein distance allow for temporal misalignments and sequence gaps, while also satisfying metricity and consistency. Thus, the method we propose in the next sections can be used for time-series data in conjunction with the discrete Fréchet distance or the ERP distance, whereas for string sequences our method can be used in conjunction with the Levenshtein distance.

## 6. INDEXING USING A REFERENCE NET

Based on the previous discussion, if a distance function is consistent, we can quickly identify a shortlist of possible similar subsequences by matching segments of the query with fixed-length window segments from the database. A simple approach is to use a linear scan for that, but this can be very expensive, especially for large databases. Therefore, it is important to use an index structure to speed up this operation.

Assuming that the distance function is a metric, we can use one of the existing index structures for metric distances. One approach is to use reference based indexing[36], as it was used in[30] for subsequence matching of DNA sequences. However, reference-based indexing has some important issues. First, the space overhead of the index can be large for some applications. Indeed, we need to use at least a few tens of references per database and this may be very costly for large databases, especially if the index has to be kept in main memory. Furthermore, the most efficient reference based methods, like the Maximum Pruning algorithm in [35], need a query sample set and a training step that can be expensive for large datasets. Therefore, we want to use a structure that has good performance, but at the same time occupies much smaller space to be stored in main memory and without the need of a training step. Another alternative is to use the Cover tree [6], which is a data structure that has linear space and answers nearest neighbor queries efficiently. Actually, under certain assumptions, the cover tree provides logarithmic query time for nearest neighbor retrieval. The main issue here is that the query performance of the cover tree for range queries may not be always good. As we show in our experiments, for some cases the performance of the cover tree can deteriorate quickly with increasing range size.

To address the issues discussed above, we propose a new generic indexing structure that is called *Reference Net*. The Reference Net can be used with any metric distance and therefore is suitable for many applications besides subsequence matching. Unlike the approaches in [36, 35], it uses much smaller space that is controllable and still provides good performance. Unlike the cover tree, every node in the structure is allowed to have multiple parents and this can improve the query performance. Furthermore, it is optimized to answer efficiently range queries for different range values that can also be controlled. Therefore, the Reference Net is a good fit for our framework.

The reference net is a hierarchical structure as shown in Figure 1. The bottom level contains all original data in the database. The structure has $r$ levels that go from 0 to $r - 1$ and in each level, other than the bottom level, we maintain a list of references. The references are data from the database. The top level has only one reference. In each level $i$, we have some references that correspond to ranges with radius $\epsilon_i = \epsilon' * 2^i$. References in the same level should

1583

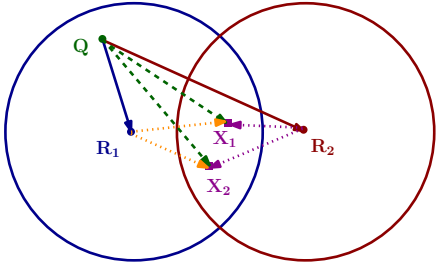

Figure 2: Difference between Net and Tree

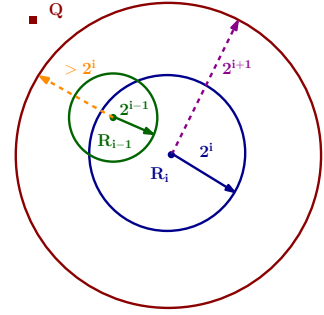

Figure 3: Intuition of Lemma 4

have a distance at least $\epsilon_i$. Let $Y_i := \{R(i,j)\}$ be the set that contains the references in level $i$. Each reference $R(i,j)$ is associated with a list $L(i,j)$ that includes references from the level below(i.e, $Y_{i-1}$). In particular, it contains references with distance less or equal to $\epsilon_i$, that is:

$$L(i,j) := \{z \in Y_{i-1} | \delta(R(i,j), z) \leqslant \epsilon_i\} \quad (1)$$

Notice that every node in the hierarchy can have multiple parents, if it is contained in multiple lists at the same level. This is one main difference with the cover tree. Another difference is that we can set the range $\epsilon'$. This helps to create a structure that better fits the application. Furthermore, this structure can answer more efficiently range queries than the cover tree. Another structure that is similar to ours is the navigating nets [19]. However, in navigating nets, the space can be more than linear. Each node in the structure has to maintain a large number of lists which makes the space and the update overhead of this structure large and the update and query algorithm more complicated.

Similar to the cover tree, the reference net has the following inclusive and exclusive properties for each level $i$:

- *Inclusive* : $\forall R(i-1, k) \in Y_{i-1}, \exists R(i,j) \in Y_i, \delta(R(i,j), R(i-1,k)) \leqslant \epsilon_i$, namely, $R(i-1,k) \in L(i,j)$.

- *Exclusive* : $\forall$ pairs of $R(i,p) \in Y_i$ and $R(i,q) \in Y_i$, $\delta(R(i,p), R(i,q)) > \epsilon_i$.

The inclusive property means that if a reference appears in the level $i-1$ it should appear in at least one reference list in the level $i$ (any reference has at least one parent in the hierarchy.) The exclusive property says that for two references to be at the same level they should be far apart (at least $\epsilon_i$).

In Figure 2 we show why it is important to have a multi-parent hierarchy and not a tree. Assume $\delta(R_1, X_i) \leqslant \varepsilon$ and $\delta(R_2, X_i) \leqslant \varepsilon$, but $X_i$ are only in the list of $R_2$. If $\delta(Q, R_2) + \varepsilon > r$ we do not know whether $\delta(Q, X_i) \leqslant r$ or not. However, if we maintain $X_i$ also in $R_1$, and $\delta(Q, R_1) + \varepsilon \leqslant r$, we know $\delta(Q, X_i) \leqslant r$ by checking $\delta(Q, R_1)$. A similar idea has also been used in the mvp-tree [7], which is another structure that is optimized for nearest neighbor and not for range queries.

Notice that we do not have to maintain empty lists or the list of a reference to itself. Indeed, according to the definition, when a reference appears in the hierarchy at the level $i$, it will appear in all levels between $i-1$ and 0. However, we just keep each reference only in the highest level. Another issue is that, depending on the data distribution and the value of $\epsilon'$, we may have many parent links from some level to the next and this may increase the space overhead. In order to keep the space linear and small, we impose a restriction on the number of lists that can contain a given reference to $num_{max}$. In most of our experiments, this was not an issue, but there are cases where this helps to keep the space overhead in check.

The advantage of the reference net compared to the reference based methods is that using a single reference we can prune much more data from the database. This is exemplified in the following lemma:

*Lemma 4.* Let $R(i,j) \in L(i,j)$. If $\delta(Q, R(i,j)) > \epsilon_{i+1}$, $\forall R(i-1,k) \in L(i,j)$, $\delta(Q, R(i-1,k)) > \epsilon_i$.

Proof: Since $R(i-1, k) \in L(i,j)$, then according to definition of reference:

$$\delta(R(i-1,k), R(i,j)) \leqslant \epsilon' * 2^i \quad (2)$$

While:

$$\delta(Q, R(i,j)) > \epsilon' * 2^{i+1} \quad (3)$$

Then according to triangular inequality:

$$\delta(Q, R(i-1,k)) > \epsilon' * 2^i \quad (4)$$

Then for any reference $R(l,k)$, $l < i$ derived from reference $R(i,j)$, $\delta(R(l,k), R(i,j)) < 2^{i+1}$. If $\delta(Q, R(i,j)) - 2^{i+1} > r$, $\delta(Q, R(l,k)) > r$. If $\delta(Q, R(i,j)) + 2^{i+1} \leqslant r$, $\delta(Q, R(l,k)) \leqslant r$. So, not only we can prune the references in one list, but we may also prune all references derived from that list. A simple example is illustrated in Figure 3.

For more details on the algorithms for insertion, deletion, and range query for reference nets see the Appendix.

## 7. SUBSEQUENCE MATCHING

Using the concepts introduced in the previous sections, the subsequence matching framework proposed in this paper consists of five steps:

1. Partition each database sequence into windows of length $\lambda/2$.

2. Build the hierarchical reference net by inserting all windows of length $\lambda/2$ from the database.

3. Extract from the query $Q$ all segments with lengths from $\lambda/2 - \lambda_0$ to $\lambda/2 + \lambda_0$.

4. For each segment extracted from $Q$, conduct a range query on the hierarchical reference net, so as to find similar windows from the database.

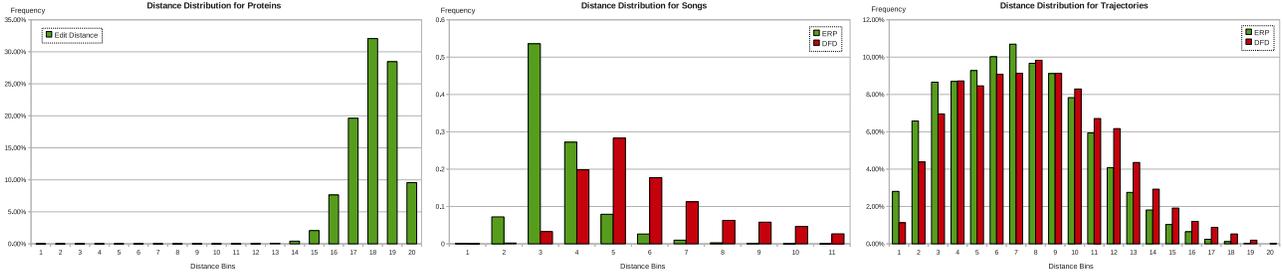

**Figure 4: Distance distribution for different data sets**

5. Using the pairings between a query segment and a database window identified in the previous step, generate candidates and identify the pairs of similar subsequences that answer the user's query.

In the next few paragraphs, we describe each of these steps.

The first two steps are offline preprocessing operations. The first step partitions each database sequence $X$ into windows of length $\lambda/2$, producing a total of $2/\lambda * |X|$ windows per sequence $X$. At the second step, we build a hierarchical reference net.

The next three steps are performed online, for each user query. Step 3 extracts from $Q$ all possible segments of lengths between $\lambda/2 - \lambda_0$ and $\lambda/2 + \lambda_0$. This produces at most $(2\lambda_0 + 1) * |Q|$ segments. At step 4 we conduct a range query for each of those segments. Step 4 outputs a set of pairs, each pair coupling a query segment with a database window of length $\lambda/2$. Note that, it is possible that many queries are executed at the same time on the index structure in a single traversal.

Given a database sequence $X$, $2/\lambda * |X|$ windows of $X$ are stored in the reference net. Thus, each query subsequence is compared to $2/\lambda * |X| * (1 - \alpha)$ windows from $X$, where $\alpha < 1$ is the pruning ratio attained using the reference net. Since there are at most $(2\lambda_0 + 1) * |Q|$ query segments, and $\lambda$ is a constant, the total number of segment pairs between $Q$ and $X$ that must be evaluated is:

$$2(2\lambda_0 + 1)/\lambda * (|X||Q|) * (1 - \alpha) \rightarrow O(|X||Q|) \quad (5)$$

In the experiments we show that the pruning ratio $\alpha$ of the proposed reference net is better than the ratios attained using cover tree and maximum variance.

The final step in our framework finds the pairs of similar subsequences that actually answer the user's query. These pairs are identified based on the pairs of subsequences from step 4. Let $SSQ_{a,b} = (Q_a, Q_{a+1}, ..., Q_b)$ and $SSX_c = (X_c, X_{c+1}, ..., X_{c+\lambda/2})$ be a pair of segments from step 4. We must identify supersequences $SQ$ of $SSQ_{a,b}$ and $SX$ of $SSX_c$ that should be included in the query results. It suffices to consider sequences $SQ$ whose start points are from $a - \lambda/2 - \lambda_0$ to $a$, and whose endpoints are between $b$ and $b + \lambda/2 + \lambda_0$. Similarly, it suffices to consider subsequences $SX$ whose starting points are between $c - \lambda/2$ and $c$, and whose endpoints are between $c + \lambda/2$ and $c + \lambda$.

As described in Section 3.2, we consider three query types. For the first type, step 5 checks all pairs of possible similar subsequences and returns all the pairs that are indeed similar. However, this query type would generate a lot of results according to the consistency property. For the second and third type, only optimal results will be returned. Also, not all pairs of possible similar subsequences need to be checked.

For query type II, i.e., Maximize $|SX| \geqslant \lambda$, Subject to $||SQ| - |SX|| \geqslant \lambda_0$ and $\delta(SX, SQ) \leqslant \varepsilon$, we have:

1. Conduct a range query with radius $\varepsilon$ (step 4.) Get a set of similar segments. If there is no similar segments, there cannot be any similar subsequences.

2. Find the longest similar subsequences: If $\langle x_i, q_j \rangle$ and $\langle x_{i+1}, q_{j+1} \rangle$ are two pairs of segments in the results, they can be concatenated. After concatenation, assume the longest sequence of segments has length $k\lambda/2$, then the longest similar subsequence has length no longer than $(k+2)\lambda/2$. Then, we start the verification from the longest sequence of segments.

If $k > 1$, at least one pair of subsequences with length at least $k\lambda/2$ will be similar. On the other hand, if $k = 1$, there may not exist similar subsequences at all.

For query type III, i.e., Minimize $\delta(SX, SQ)$, Subject to $||SQ| - |SX|| \geqslant \lambda_0$ and $\delta(SX, SQ) \leqslant \varepsilon$, we have.

1. Use binary search to find the minimal $\varepsilon$ that gives at least a pair of similar segments in step 4.

2. Find the longest similar subsequences: Conduct step (2) of query type II to get the longest similar subsequences. If we find some results, the current $\varepsilon$ is optimal.

3. If there is no similar subsequence, increase $\varepsilon$ by an increment $\epsilon_{inc}$ and find similar segments. Then redo step (2).

The increment $\epsilon_{inc}$ depends on the dataset and the distance function and can be a constant factor of the minimum pairwise distance in the dataset.

## 8. EXPERIMENTS

In this section we present an experimental evaluation of the reference net using different datasets and distance functions. The goal of this evaluation is to demonstrate that the proposed approach works efficiently for a number of diverse datasets and distance functions.

The first dataset is a protein sequence dataset [1] (*PROTEINS*). Proteins are strings with an alphabet of 20 and the distance function is the Levenshtein (Edit) distance.

---
[1] http://www.ebi.ac.uk/uniprot/

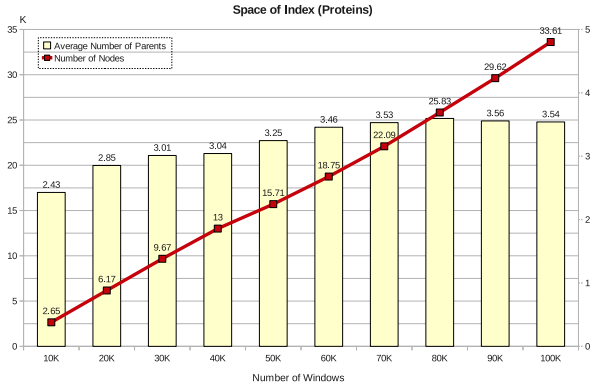

**Figure 5: Space overhead for** *PROTEINS*

The protein sequences are partitioned into $100K$ total windows of size $l = 20$. We also use two different time series datasets. One is the *SONGS* dataset, where we use sequences of pitches of songs as time series from a collection of songs [5] and the other is a trajectory (*TRAJ*) dataset that was created using video sequences in a parking lot [37]. The *SONGS* dataset contains up to $20K$ windows and the *TRAJ* dataset up to $100k$ windows both of size $l = 20$. For the time series datasets we used two distance functions: the Discrete Frechét Distance (DFD) and the ERP distance.

First, we show the distance distributions of these datasets and distance functions in Figure 4. It is interesting to note that for the *SONGS* dataset, since the pitch values range between 0 and 11, the DFD distribution is very skewed and most of the distances are between 2 and 5. On the other hand, the ERP distance on the same dataset gives a distance distribution that is more spread out. As we will see later this can affect the index performance.

## 8.1 Space Overhead of Reference Net

Here we present the space overhead of the reference net index for each dataset. In all our experiments we used a default value for $\epsilon' = 1$. In Figure 5, we plot the space consumption of the reference net for the *PROTEINS* dataset under the Levenshtein distance. We show the number of nodes in the index in thousands, when the number of inserted windows ranges from $10K$ to $100K$. As we can see, this increases linearly with the number of inserted windows. We plot also the average size of each reference list in the net, which is in general below 4. Note that the average size of each list is actually the average number of parents for each node. Therefore, the size of the reference net is about three to four times more than the size of the cover tree where each node has only one parent. Finally, the total size of the index for $100K$ windows is about $2.9MBytes$.

In Figure 6, we show the results for the *SONGS* dataset using the two different distance functions DFD and ERP. In the first plot we show the number of reference lists (top three lines) and the size of the index in $MBytes$ for different number of windows ranging from $1K$ to $20K$. Recall that the DFD distance creates a very skewed distribution for this dataset. Therefore, the number of references as well the size of the index is much larger than using the ERP. The reason is explained in the next plot, where we show the average number of parents per window. As we can see, inserting more and more windows increases the size of the reference lists for DFD, since most new windows have small distance with many other existing ones and the number of parents increases. We have to mention that here we did not restrict the size of the number of parents yet. On the other hand, the ERP distance creates a more wide-spread distance distribution and the average number of parents remains small. Then, we impose a constraint and limit the maximum number of lists that a given window can appear to $num_{max} = 5$ and we call this DFD-5. Notice that the average number of parents per window is now below 5. The reason is that all windows that can have more than 5 parents in the unconstrained case are limited to the have exactly 5. As we can see, the size of the index now decreases and is similar to the index created with the ERP distance.

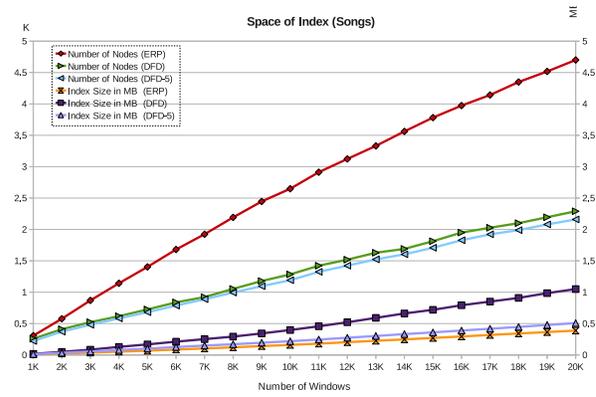

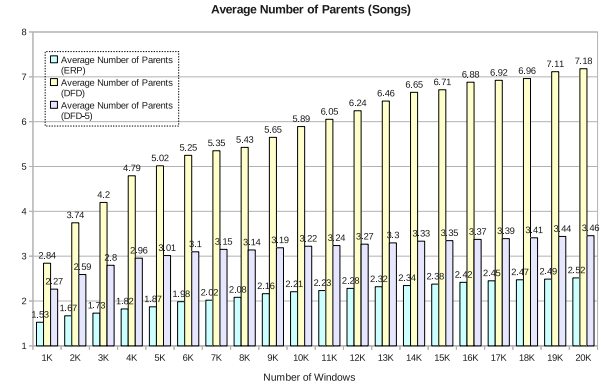

**Figure 6: Space overhead for** *SONGS*

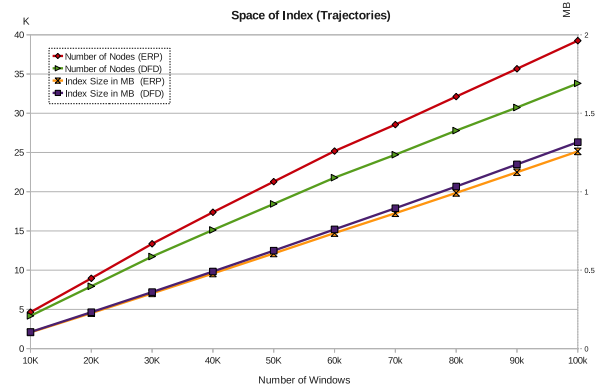

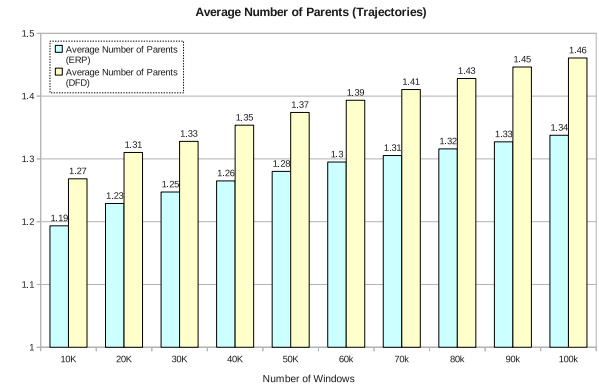

**Figure 7: Space overhead for** *TRAJ*

Finally, in Figure 7, we show the results for the *TRAJ* dataset. Since the variance of the distance distribution is now higher for both distance functions, the reference net has small average number of parents per window and the size of the index is small. Actually, in that case the size of the index is less than twice the size of the cover tree.

## 8.2 Query Evaluation

Here we present the query performance of reference net (RN) compared against the cover tree (CT) and the reference based indexing that uses similar or larger space. For the reference based method we use the Maximum Variance (MV) approach to select references [36]. The main reason is the we did not have enough training data for the Maximum Pruning (MP) approach and actually it performed similarly with the MV for the queries that we used. The MV method is also much faster to compute.

In Figure 8 we show the performance of all the indices on the *PROTEINS* for range queries with different sizes. Here we show the percentage of the distance computations that we need to perform using the different indices against the naive solution where we compute the distance of the query with each window in the database. We can verify that the reference net performs better than the cover tree as expected. Furthermore, the MV-5, which has the same space requirement as the reference net, performs much worse. Increasing the size of the MV method by a factor of 10 (MV-50), helps to improve the performance for very small ranges, but when the range size increases a little bit (becomes 10% of the maximum distance) the performance of MV-50 becomes similar to the reference net and then becomes worse. Notice that the maximum distance is 20 and therefore the 10% means a distance of 2 in our case.

In the Figure 9, we see the results for the *SONGS* dataset and the DFD distance. Notice that the RN-5 which is the reference net with the constraint that $num_{max} = 5$ has similar performance with the unconstrained reference net. Again the performance is better than the cover tree and the MV with similar space. We got similar results with the ERP distance for this dataset.

In Figure 10, we show the performance for the *TRAJ* dataset and the ERP distance. In addition to the percentage of distance computations versus the naive solution we show in the plot the pairwise distance distribution for this dataset and distance function. In particular, for each query range we show the distribution of the pairs of sequences that have this distance. It is interesting to note that the performance of the index methods follow the distribution of the distance values. Furthermore, the cover tree and the reference net have similar performance since they have similar space and structure. However, they perform much better than the MV-20 methods which has 10 times more space. We get similar results for the DFD distance as we can see in Figure 11.

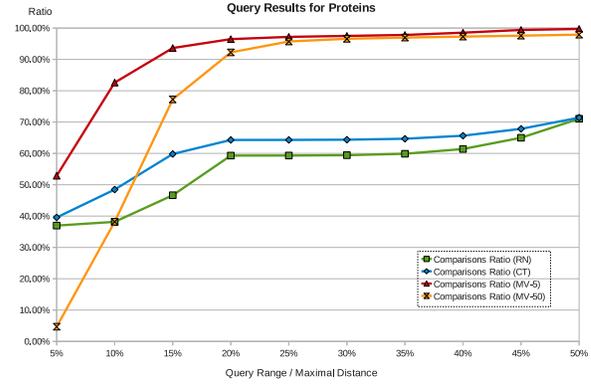

Figure 8: Query performance for *PROTEINS*

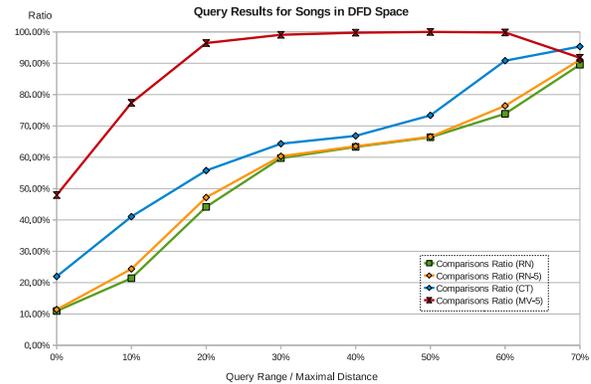

Figure 9: Query performance for *SONGS* and DFD

Overall, the reference net performs better than the cover tree and much better than the MV approach when they use the same space. Actually, sometimes it performs better than the MV method even when we use 10 time more space than the reference net.

Finally, in Figure 12, we report some results on the number of unique windows that match at least one segment of the query for the *PROTEINS* dataset. We generated random queries of size similar to the smallest proteins in the dataset and we run a number of queries for different values of $\epsilon$. As expected, the number of unique windows in the database that have a match with the query follows the distribution of the distances. Notice that the maximum distance is 20 and therefore, when we set $\epsilon = 20$ we get the full database back. A more interesting result is the number of *consecutive windows* (at least two consecutive windows) as a percentage of the total number of windows. As we can see, this number is much smaller than the number of unique single matching windows. Therefore, for answering the query Type II, we will start first from the consecutive windows and if we succeed, we may not need to check any other matching windows that are not consecutive. This shows that for more interesting query types (II and III), we may have to check a small number of candidate matches and we can perform the refinement step very efficiently depending on the dataset and the query.

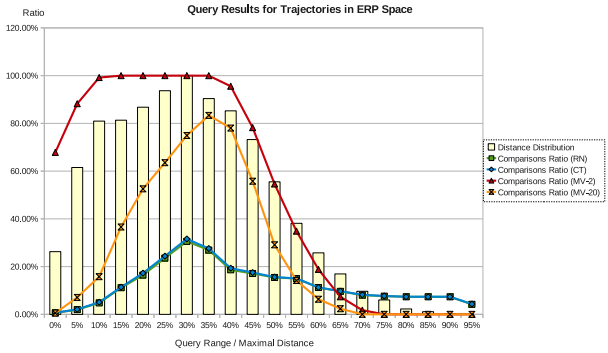

**Figure 10: Query performance for $TRAJ$ and ERP**

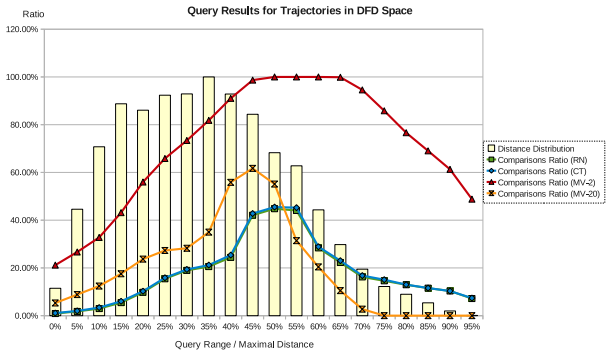

**Figure 11: Query for $TRAJ$ and DFD**

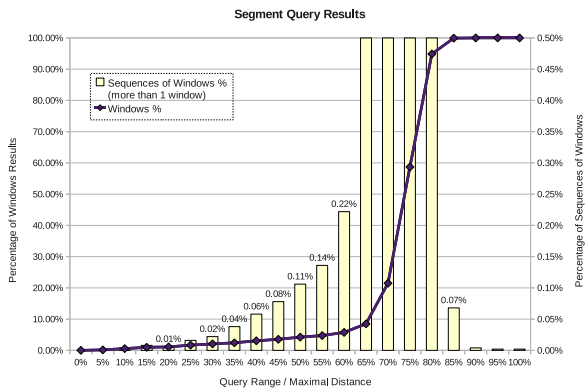

**Figure 12: Query results for $PROTEINS$-$10K$**

## 9. CONCLUSIONS

We have presented a novel method for efficient subsequence matching in string and time series databases. The key difference of the proposed method from existing approaches is that our method can work with a variety of distance measures. As shown in the paper, the only restrictions that our method places on a distance measure is that the distance should be metric and consistent. Actually, it is important to note that the pruning method of Section 5 only requires consistency, but not metricity. We show that the consistency property is obeyed by most popular distance measures that have been proposed for sequences, and thus requiring this property is a fairly mild restriction.

We have also presented a generic metric indexing structure, namely hierarchical reference net, which can be used to further improve the efficiency of our method. We show that, compared to alternative indexing methods such as the cover tree and reference-based indexing, for comparable space costs the reference net is faster than the alternatives. Overall, our experiments demonstrate the ability of our method to be applied to diverse data and diverse distance measures.

## 10. ACKNOWLEDGMENTS

This work has been partially supported by NSF grants IIS-0812309, IIS-0812601, IIS-1055062, CNS-0923494, and CNS-1035913.

# APPENDIX

## A. ALGORITHMS FOR REFERENCE NET

In this Appendix, we present algorithms to (i) insert a single object into a reference net structure, (ii) delete an object from the structure, and (iii) run a range query using a reference net.

### A.1 Insertion

The insertion algorithm starts from the top level of the hierarchy of the reference net. When inserting a new object $X$, initially, the set of candidate parents, $C$, includes only the root reference and $i$ is the level where $\epsilon' * 2^{i-1} < \delta(root, X) \leqslant \epsilon' * 2^i$. Then, the algorithm keeps updating the candidate parents $C$ and level $i$ until there is no parent in $C$. With this approach, the algorithm allows to insert one object into several lists at the same level. Insertion is illustrated in algorithm 1.

**Algorithm 1:** Insert(X)

**Input**: An object $X$, initial set of candidate parents $C$, initial $i$
1. **while** $\exists\ R(i,j) \in C$ **do**
2.     Find $L(i+1,k)$, where $R(i,j) \in L(i+1,k)$;
3.     **forall the** $S \in L(i,j) \bigcup L(i+1,k)$ **do**
4.         Compute $i'$, $\epsilon' * 2^{i'-1} < \delta(S, X) \leqslant \epsilon' * 2^{i'}$ ;
5.         Update $i = min\{i'\}$ ;
6.         Update $C = \{S \mid \delta(S, X) \leqslant \epsilon' * 2^{min\{i'\}}\}$ ;
7.     **end**
8. **end**
9. Insert $X$ to $L(i*, j*)$ where $R(i*, j*) \in$ last $C$;

Notice that the insertion algorithm jumps to the lowest possible level in each update. The reason is the following: if the inserted object is in the list $L(i,j)$, it cannot belong to any list $L(i',j')$ in a higher level $(i' > i)$. Hence, we try to go to the lowest possible level in order to generate a small number candidate parents.

### A.2 Deletion

The deletion algorithm consists of two phases: First, we find the lists that the object to be deleted, $X$, belongs to and we remove it. Second, we handle all the objects in $X$'s list. Similar to the insertion algorithm, the deletion algorithm also runs from the top level of the reference net. Notice that, the objects in the $X$'s list may have to be re-distributed. If an object in this list already belongs to the list of another reference object in the same level of $X$, we do nothing. Otherwise, we need to insert this object somewhere. Therefore, we insert it into the lists that $X$ was in.

**Algorithm 2:** Delete(X)

**Input**: An object $X$
1. Find $X = R(i,j)$;
2. Remove $X$ from every $L(i+1, j_0)$;
3. **forall the** $R(i-1, k) \in L(i, j)$ **do**
4.     **if** $\nexists\ j'$, $R(i-1,k) \in L(i, j')$ **then**
5.         Re-Insert $R(i-1, k)$;
6.     **end**
7. **end**

### A.3 Range Query

The range query takes a query $Q$ and a distance $\varepsilon$, and finds all object $X$, where $\delta(Q, X) \leqslant \varepsilon$. We use lemma 4 and the triangular inequality to prune objects in each level. If $\delta(R(i,j), Q) + \epsilon' * 2^i \leqslant \varepsilon$, for every object $X$ in $L(i,j)$, $\delta(Q, X) \leqslant \varepsilon$. If $\delta(R(i,j), Q) - \epsilon' * 2^i > \varepsilon$, for every object $X$ in $L(i,j)$, $\delta(Q, X) > \varepsilon$. If $\delta(R(i,j), Q) + \epsilon' * 2^{i+1} \leqslant \varepsilon$, for every object $X$ derived from $R(i,j)$, $\delta(Q, X) \leqslant \varepsilon$. If $\delta(R(i,j), Q) - \epsilon' * 2^{i+1} > \varepsilon$, for every object $X$ derived from $R(i,j)$, $\delta(Q, X) > \varepsilon$. We actually maintain two sets: $C$ and $P$. $C$ stores all objects which are definitely in the results of the query. $P$ includes all objects that are not in the result.

**Algorithm 3:** Range Query(Q, $\varepsilon$)

**Input**: A query $Q$ and a distance $\varepsilon$
1. **foreach** $R(i,j) \in Y_i$, $i$ from $r-1$ to $0$ **do**
2.     **if** $R(i,j) \notin C \bigcup P$ **then**
3.         Compute $d = \delta(R(i,j), Q)$;
4.         **if** $d + \epsilon' * 2^{i+1} \leqslant \varepsilon$ **then**
5.             Insert all $X$ derived from $R(i,j)$ to $C$;
6.         **else if** $d + \epsilon' * 2^i \leqslant \varepsilon$ **then**
7.             Insert all $X \in L(i,j)$ to $C$;
8.         **end**
9.         **if** $d - \epsilon' * 2^{i+1} > \varepsilon$ **then**
10.            Insert all $X$ derived from $R(i,j)$ to $P$;
11.         **else if** $d - \epsilon' * 2^i > \varepsilon$ **then**
12.             Insert all $X \in L(i,j)$ to $P$;
13.         **end**
14.     **end**
15. **end**
16. Expand all objects in $C$;

Note that sometimes $\epsilon' * 2^i < \delta(R(i,j), Q) - \varepsilon \leqslant \epsilon' * 2^{i+1}$ holds. And for some $R(i-1, k) \in L(i,j)$, $\delta(R(i-1,k), Q) - \varepsilon > \epsilon' * 2^{i-1}$. If we prune all $R(i-1,k) \in L(i,j)$ at level $i$, then we cannot prune all $X$ derived from $R(i-1,k)$ at level $i-1$.